\documentclass[10pt,journal,compsoc]{IEEEtran}
\usepackage{graphicx}
\usepackage{multirow}
\usepackage{dblfloatfix} 
\usepackage[dvipsnames]{xcolor}
\usepackage{soul}
\usepackage{color}
\usepackage{url}

\usepackage{boldline}
\usepackage{xcolor,colortbl}
\usepackage{tikz}
\def\checkmark{\tikz\fill[scale=0.4](0,.35) -- (.25,0) -- (1,.7) -- (.25,.15) -- cycle;}
\usepackage{pifont}

\newcommand{\xmark}{\text{\ding{55}}}

\definecolor{Gray}{gray}{0.85}
\definecolor{LightCyan}{gray}{0.85}

\newcolumntype{g}{>{\columncolor{Gray}}c}
\newcommand*{\SL}[1]{\textcolor{red}{#1}}

\usepackage[separate-uncertainty = true,multi-part-units = repeat]{siunitx}
\usepackage{array,amsfonts,amsmath}
\usepackage[font=small]{caption}
\usepackage{subcaption}

\usepackage[english]{babel}
\usepackage{lipsum}
\usepackage[pscoord]{eso-pic}

\ifCLASSOPTIONcompsoc
  \usepackage[nocompress]{cite}
\else
  \usepackage{cite}
\fi
\ifCLASSINFOpdf
\else
\fi
\newcommand{\placetextbox}[3]{
  \setbox0=\hbox{#3}
  \AddToShipoutPictureFG*{
    \put(\LenToUnit{#1\paperwidth},\LenToUnit{#2\paperheight}){\vtop{{\null}\makebox[0pt][c]{#3}}}%
  }%
}%

\placetextbox{0.5}{1}{\scriptsize{\SL{This work has been accepted in IEEE Transactions on Affective Computing for possible publication. Copyright may be transferred without notice and this version may no longer be accessible.}}}

\begin{document}
\title{Multi-Task Semi-Supervised Adversarial Autoencoding for Speech Emotion Recognition}

\author{Siddique Latif, Rajib Rana, Sara Khalifa, Raja Jurdak, Julien Epps, and Bj\"{o}rn W.\ Schuller,~\IEEEmembership{Fellow,~IEEE}
\IEEEcompsocitemizethanks{
\IEEEcompsocthanksitem  S.\ Latif is affiliated with USQ, Australia and Distributed Sensing Systems Group, Data61, CSIRO Australia. 
\IEEEcompsocthanksitem R. Rana is with University of Southern Queensland (USQ), Australia.
\IEEEcompsocthanksitem  S.\ Khalifa is affiliated with Distributed Sensing Systems Group, Data61, CSIRO Australia.
\IEEEcompsocthanksitem  R.\ Jurdak is affiliated with  Queensland University of Technology (QUT), Australia.
\IEEEcompsocthanksitem  J.\ Epps is affiliated with University of New South Wales (UNSW), Australia.
\IEEEcompsocthanksitem  B.\ Schuller is affiliated with GLAM -- the Group on Language, Audio, and Music, Imperial College London, UK, and the ZD.B Chair of Embedded Intelligence for Health Care and Wellbeing, University of Augsburg, Germany.

Corresponding E-mail: siddique.latif@usq.edu.au}}


\IEEEtitleabstractindextext{%
\begin{abstract}

Inspite the emerging importance of Speech Emotion Recognition (SER), the state-of-the-art accuracy is quite low and needs improvement to make commercial applications of SER viable. A key underlying reason for the low accuracy is the scarcity of emotion datasets, which is a challenge for developing any robust machine learning model in general.  In this paper, we propose a solution to this problem: a multi-task learning framework that uses auxiliary tasks for which data is abundantly available. We show that utilisation of this additional data can improve the primary task of SER for which only limited labelled data is available.  In particular, we use gender identifications and speaker recognition as auxiliary tasks, which allow the use of very large datasets, e.\,g., speaker classification datasets. To maximise the benefit of multi-task learning, we further use an adversarial autoencoder (AAE) within our framework, which has a strong capability to learn powerful and discriminative features. Furthermore, the unsupervised AAE in combination with the supervised classification networks enables semi-supervised learning which  incorporates a discriminative component in the AAE unsupervised training pipeline. This semi-supervised learning essentially helps to improve generalisation of our framework and thus leads to improvements in SER performance. The proposed model is rigorously evaluated for categorical and dimensional emotion, and cross-corpus scenarios. Experimental results demonstrate that the proposed model achieves state-of-the-art performance on two publicly available datasets.

\end{abstract}

}

\maketitle

\IEEEdisplaynontitleabstractindextext
\IEEEpeerreviewmaketitle

\IEEEraisesectionheading{\section{Introduction}\label{sec:introduction}}
\IEEEPARstart{S}{peech} Emotion Recognition (SER) is an emerging area of research. Since speech is a major form of affect display \cite{trower2013social}, the success of SER will redefine human-computer interactions, enabling, for example, effective service delivery in many sectors. Call centres now track customers' emotions for better service delivery~\cite{burkhardt2006detecting}. Speech based diagnostic systems are being developed for diagnosis of depression \cite{huang2019speech}, distress \cite{rana2019automated}, and monitoring of mood states for bipolar patients \cite{huang2018detecting}. Many other applications including media retrieval systems \cite{merler2018automatic}, smart cars \cite{vogel2018emotion}, and forensic sciences \cite{roberts2012forensic} also aim to improve their performances by utilising SER techniques.
 
Human emotions in speech are complex to model due to dependency of speech on many factors including speaker~\cite{ding2012speaker}, gender~\cite{vogt2006improving}, age~\cite{mill2009age}, culture~\cite{latif2018cross}, dialect~\cite{laukka2014evidence}, and among others. Researchers have explored many methods including classical models, such as  hidden Markov models, support vector classification, and deep neural networks (DNNs) for speech emotion recognition, wherein DNN models have usually demonstrated better performance compared to the classical models\cite{schuller2018speech,latif2019direct}. Currently, the popularity of DNN models for speech emotion recognition is seeing a steep rise.

DNN models that have been successful for speech emotion recognition include deep belief networks (DBN) \cite{kahou2013combining,Latif2018}, convolutional neural networks (CNN) \cite{mao2014learning,Neumann2017} and long short term memory (LSTM) networks  \cite{Latif2018v,mirsamadi2017automatic,wollmer2012analyzing}. The majority of the above research presents techniques to predict speech emotion using single task (emotion recognition) training. These techniques, however, ignore a potentially rich source of information available in speech (e.\,g., information about the speaker, gender, etc.) that can be utilised for achieving generalisation and improvement in the performance \cite{zhang2019attention}. To achieve generalisation, most existing studies tend to validate/tune models using diverse datasets \cite{Latif2018,latif2018cross}. However, standard benchmark datasets are very scarce, and most problematically, they are of smaller sizes, which creates massive roadblocks in achieving generalisation in SER systems \cite{zhang2019attention}.

An alternative and effective approach to increase the generalisation of SER models is multi-task learning (MTL) \cite{caruana1997multitask}, which simultaneously solves relevant auxiliary tasks along with the primary task. In MTL, models are better regularised to uncover the common high-level discriminative representations. MTL has been widely applied to various speech and natural language processing related problems \cite{li2011multi,collobert2008unified}. In SER, MTL has shown good performance for fully supervised deep learning (DL) models \cite{kim2017towards,parthasarathy2017jointly,zhang2019attention}. Most of these approaches jointly learn different emotional attributes to improve both performance and generalisation~\cite{Lotfian2018,parthasarathy2017jointly,xia2017multi}. Often, researchers use categorical emotion as a primary task and dimensional emotion as an auxiliary task. Discrete/categorical theories of emotions encompass a small set of distinct emotions. The foundation of these theories is that different emotions are associated with distinctive patterns of triggers, behavioural expression, and unique subjective experiences \cite{hudlicka2017computational}. Only core emotions including joy, sadness, fear, anger, and disgust \cite{ekman1992argument} are included in these theories. On the contrary, the foundation of the dimensional models of emotions is that a common and interconnected neurophysiological system generates all affective states \cite{posner2005circumplex}.  Generally, these models define human emotions using a two-dimensional space having valence in one dimension and activation or intensity in the other dimension. 
To use these dimensional emotions as secondary tasks, annotation is important. However, the meta labels for these emotional attributes are scarcely available. Recently, it has been shown in computer vision that the performance of primary tasks with constrained data can be significantly enhanced by using larger data for the auxiliary tasks \cite{liu2019exploiting,liu2018leveraging}. Inspired by this idea, in this study, we aim to build models that can effectively utilise auxiliary tasks with a large quantity of available data in order to improve the performance of the primary task. We use emotion recognition as our primary task and select gender and speaker recognition as auxiliary tasks  to include larger datasets. 


Within our MTL framework, we further utilise generative adversarial models due to their exceptional ability to learn powerful and discriminative features\cite{sahu2018adversarial}. In particular, we use adversarial autoencoder (AAE) \cite{makhzani2015adversarial}, which fundamentally aims to learn representation of data in an unsupervised way. However, by combining AAE with the supervised classification networks we enable semi-supervised learning for AAE. This essentially incorporates the discriminative component (from the supervised classification) in the training pipeline of unsupervised learning to influence the latent representation of AAE and by makes it suitable for semi-supervised emotion classification.





To show the advantage of our proposed MTL framework, we evaluate it comprehensively on two large and widely used emotional databases: The interactive emotional dyadic motion capture (IEMOCAP) \cite{busso2008iemocap} and MSP-IMPROV \cite{busso2017msp}. We compare the performance of our proposed framework with that of recent studies, and  also with popular models like CNNs, and an autoencoder based semi-supervised model. The comparative results show that, for categorical, dimensional, and cross-corpus emotion classification, we achieve the improved results, which attests to the strong generalisation power of the proposed framework. 

\section{Related Work}
Our framework utilises multi-task learning for SER. It also uses semi supervised learning while employing  adversarial encoding, where the classification is done through CNNs. We therefore cover these four aspects in our literature review.
\subsection{Landscape of Multi-task Learning for SER}
Multi-task learning (MTL) has been successful for simultaneously modelling multiple related tasks utilising shared representation  \cite{ben2003exploiting,baxter2000model}. It aims to improve generalisation by learning the similarities as well as the differences among the given tasks from the training data \cite{caruana1997multitask}. The conventional methodology to optimise a machine learning model for one task at a time ignores potentially rich information in the training signal \cite{zhang2019attention}. Such information can be effectively utilised for auxiliary tasks to improve generalisation and performance of a system. Several MTL approaches \cite{li2014heterogeneous,zhang2014facial,farabet2013learning} have been widely used for solving problems in computer vision. The primary reason to use MTL in vision is that images can provide information related to different tasks, and simultaneously learning these correlated tasks can boost the performance of each individual task \cite{chen2014joint,zhu2012face}. For example, face detection, gender recognition, and pose estimation can be simultaneously performed using a deep CNN \cite{ranjan2019hyperface}. 

Similarly to images, speech is another such modality that can provide information for various tasks including speaker, gender, and emotion identification. Researchers have started to investigate the effectiveness of MTL for improving the performance of speech emotion recognition \cite{kim2017towards,le2017discretized,ma2018speech}. Eyben et al.\ \cite{eyben2012multitask} were first to use MTL in SER and they showed that training the model with multiple targets helps to improve the performance compared to single target training. Prthasarathy and Busso \cite{parthasarathy2017jointly} proposed a DNN based model to jointly learn the arousal, dominance, and valence value of a given utterance. The authors demonstrated that joint learning of these emotional attributes significantly enhances the performance of a model compared to single task learning (STL). Similarly, Ma et al.\ \cite{ma2018speech} used a multi-task attention-based DNN for SER and found that, by sharing the information among tasks, a high performance can be achieved.  Xia et al.\ \cite{xia2017multi} proposed a DBN based model for MTL and utilised activation and valence information for speech emotion recognition. They illustrated that the utilisation of additional information in the MTL setup can improve the performance of their model by considering the categorical emotion label as the primary task, and activation and valence information as secondary tasks. 
Similarly, Lotfian et al.\ \cite{Lotfian2018} used a DNN for jointly learning primary and secondary emotions. They showed that the classification performance of the primary task (categorical emotions) is significantly improved by considering secondary emotions (other emotional
classes perceived by the evaluators) in the model. In another study, Chang et al.\ \cite{chang2017learning} used a generative adversarial network (GAN) for MTL with valence classification as primary and activation classification as a secondary task. In addition, the authors also introduced unlabelled data from the AMI corpus \cite{carletta2005ami} (a multi-modal data set consisting of 100 hours of meeting recordings) to train generator and discriminator components of a GAN and showed that the performance of the classifier can be improved by using task-unrelated speech data in an unsupervised way. 

Another stream of research in SER---instead of using different emotional attributes as auxiliary tasks---has utilised other available attributes, such as speaker identity and gender to improve the performance of SER \cite{tao2018advanced}. For instance, Kim et al.\ \cite{kim2017towards} used gender and naturalness (natural or acted corpus) recognition as auxiliary tasks to improve the performance of emotion recognition using different emotional databases. Zhang et al.\ \cite{zhang2017cross} used an MTL approach to investigate the influence of the domain (whether the expression is spoken or sung), corpus, and gender on cross-corpus emotion recognition systems. The authors used six different emotional databases and showed that the performance of a cross-corpus SER system increases with the rising number of emotional corpora used for training. Based on these results, they also showed that effective modelling of cross-corpus emotion recognition requires the understanding of emotional changes as a function of non-emotional factors.


\begin{table*}[]
\centering
 \caption{Summary of comparative analysis of our paper with that of the existing literature.}
\label{table:comparison}
\begin{tabular}{|l|c|c|c|c|>{\columncolor[HTML]{EFEFEF}}c |} \hline
 & \multicolumn{2}{c|}{Auxiliary Tasks for MTL} & &   & \cellcolor[HTML]{EFEFEF}   \\ \cline{2-3}
\multirow{-2}{*}{Paper/Author  (Year)} 
& \begin{tabular}[c]{@{}c@{}}Emotional \\ Attributes\end{tabular}
& \begin{tabular}[c]{@{}c@{}}Non-Emotional \\ Attributes\end{tabular} 
& \multirow{-2}{*}{\begin{tabular}[c]{@{}c@{}}Adversarial \\ Learning\end{tabular}} 
& \multirow{-2}{*}{\begin{tabular}[c]{@{}c@{}}Semi-Supervised \\  Learning\end{tabular}} 
& \multirow{-2}{*}{\cellcolor[HTML]{EFEFEF}\begin{tabular}[c]{@{}c@{}}Additional Data for \\ Auxiliary Tasks\end{tabular}} \\ \hline

\begin{tabular}[l]{@{}l@{}}Prthasarathy and \\ Busso \cite{parthasarathy2017jointly} (2017)\end{tabular} &\checkmark{}  & \xmark{} &\xmark{}     &\xmark{}   &  \xmark{} \\ \hline
 Xia et al.\ \cite{xia2017multi} (2017) &\checkmark{}  & \xmark{} &\xmark{}     &\xmark{}   &  \xmark{} \\ \hline

Chang et al.\ \cite{chang2017learning} (2017) &\checkmark{}  & \xmark{} &\checkmark{}     &\checkmark{}  &  \xmark{} \\ \hline
Lotfian et al.\ \cite{Lotfian2018} (2018) &\checkmark{}  & \xmark{} &\xmark{}     &\xmark{}  &  \xmark{} \\ \hline
Tao et al.\ \cite{tao2018advanced} (2018)  &\xmark{}  & \checkmark{} &\xmark{}     &\xmark{}  &  \xmark{} \\ \hline
Zhang et al.\ \cite{zhang2018leveraging} (2018) &\xmark{}  & \xmark{} &\xmark{}     & \checkmark{} &  \xmark{} \\ \hline
Deng et al.\ \cite{deng2018semisupervised} (2018) &\xmark{}  & \xmark{} &\xmark{}     & \checkmark{} &  \xmark{} \\ \hline
Huang et al.\ \cite{huang2018speech} (2018) &\xmark{}  & \xmark{} &\xmark{}     & \checkmark{} &  \xmark{} \\ \hline
Sahu et al.\ \cite{sahu2018adversarial} (2018) &\xmark{}  & \xmark{} &\checkmark{}   &\xmark{}  &  \xmark{} \\ \hline
Tao et al.\ \cite{Tao2019} (2019) &\xmark{}  & \xmark{} &\xmark{}  &\checkmark{}   &  \xmark{} \\ \hline
\begin{tabular}[l]{@{}l@{}}Prthasarathy and \\ Busso \cite{parthasarathy2019semi} (2019)\end{tabular} &\checkmark{}  & \xmark{} &\xmark{}  &\checkmark{}   &  \xmark{} \\ \hline
{\bf Our Paper (2019)} &\xmark{}  & \checkmark{}  &\checkmark{}   &\checkmark{}   &  \checkmark{} \\ \hline

\end{tabular}
\end{table*}

Both streams of research mentioned above conform to the fact that MTL approaches can improve the SER performance compared with STL. While the first stream shows that choosing emotional attributes as auxiliary tasks leads to improved performance of SER for the primary task, the second stream shows that it is also possible to choose non-emotional attributes of speech as a secondary task, and performance of SER as a primary task can be improved.  Our approach is motivated by the second stream as it provides the opportunity to utilise abundantly available non-emotional datasets. Precisely, we consider using abundantly available non-emotional speech corpora to indirectly improve the performance of SER by directly improving the performance of the auxiliary tasks, which has not been widely studied in the existing literature. In~\cite{chang2017learning}, the authors used additional data, however, unlike them, we use additional data for auxiliary tasks. Also, unlike them, we backpropagate AE reconstruction loss in addition to backpropagating classification, generator, and discriminator losses. Note that, as our training uses both labelled and unlabelled emotion data, we therefore,  introduce semi-supervised learning in MTL. In the next section, we cover studies using semi-supervised learning for SER. 

\subsection{Landscape of Semi-Supervised Learning for SER}
A number of studies have considered semi-supervised learning for SER. Huang et al.\ \cite{Huang} introduced semi-supervised CNN for learning affect-salient features and reported superior performance on four public emotional speech databases: the Surrey Audio-Visual Expressed Emotion (SAVEE) database \cite{jackson2014surrey}, the Berlin Emotional Database (Emo-DB) \cite{burkhardt2005database}, the Danish Emotional Speech database (DES) \cite{DES}, and the Mandarin Emotional Speech database (MES) \cite{fu2008speaker}. The authors used CNN in an unsupervised way to learn general features and then fine-tuned the model for emotion recognition. Zhang et al.\ \cite{zhang2018leveraging} proposed a collaborative semi-supervised learning technique that can correct mislabelled samples by re-evaluating the automatically labelled samples in learning iterations of the model. 
They also used different models including SVMs and RNNs and multiple modalities (audio and video) to improve the performance by simultaneously minimising the joint entropy. Recently, researchers further studied ladder network-based semi-supervised methods for SER \cite{huang2018speech,Tao2019,parthasarathy2019semi,parthasarathy2018ladder} and have shown superior results over supervised methods. A ladder network is an unsupervised denoising autoencoder that is trained along with a supervised classification or regression task.  Deng et al.\ \cite{deng2018semisupervised} proposed a framework for SER by combining an autoencoder and a classifier. Their work is based on a discriminative Restricted Boltzmann Machine (RBM) \cite{larochelle2012learning}, which considers unlabelled samples as an extra garbage class in the classification problem. Our study differs from previous studies by simultaneously training an adversarial autoencoder with multi-task classifiers and utilising the additional unlabelled emotional data for auxiliary tasks to improve SER performance. Joint optimisation of the sum of multi-task supervised and unsupervised cost functions is an important contribution leading to more discriminative SER models. In the next section, we focus on the existing studies that utilise adversarial autoencoders (AAE) for SER and highlight the difference with our work.

\subsection{Landscape of Adversarial Autoencoders (AAE) for SER}

Autoencoders are unsupervised learning models that have been successfully utilised in the field of SER. They are very powerful in learning salient representations that lead to a notable improvement in SER performance \cite{deng2013sparse,Latif2018v}. Adversarial autoencoders (AAEs) are probabilistic models~\cite{makhzani2015adversarial} that turn an autoencoder into a generative model. This has increased the popularity of AAEs in learning more descriptive features compared to conventional AEs and even compared to variational autoencoders (VAEs) \cite{makhzani2018unsupervised}. In~\cite{sahu2018adversarial}, AAEs have been used in SER for encoding high dimensional feature representations into compressed space and for the generation of speech samples. The authors found that the latent code learnt by AAE preserves class discriminability that is very crucial for speech emotion classification. However, most SER studies utilised AE networks to perform feature learning and then classification was performed separately (e.\,g.,~\cite{sahu2018adversarial,deng2014linked}). However, it has been shown in that AAEs can be exploited in semi-supervised way to improve the classification performance \cite{makhzani2015adversarial}. Therefore, we proposed a self-sufficient semi-supervised structure that can performs both feature representation learning and classification learning by jointly minimising reconstruction error and the sum of multi-task classification errors. 

\subsection{Landscape of CNNs for SER}
Convolutional neural networks (CNNs) are one of the most popular deep learning models that have demonstrated great success in various research fields including object recognition \cite{krizhevsky2012imagenet}, handwriting recognition \cite{lecun1990handwritten}, face recognition \cite{lawrence1997face}, natural language processing (NLP) \cite{zhang2015character}, and speech recognition \cite{xiong2018microsoft}. CNNs overcome the scalability problem of standard neural networks by allowing the multiple regions of the input to share the same weights \cite{latif2019direct}. Generally, CNNs consist of three building blocks: convolutional layers, pooling layers, and fully connected layers. Convolutional layers in CNNs perform a convolution operation to compute feature maps, which are then sub-sampled using pooling layers. Finally, fully connected layers are used to transform the features into a more discriminative space for target prediction. In SER, CNNs have been widely used to learn salient features \cite{huang2014speech,mao2014learning,liu2018learning}, also directly for classification \cite{neumann2017attentive}. Studies \cite{trigeorgis2016adieu,lim2016speech,zhao2019speech,etienne2018cnn+} also presented  CNNs in combination with LSTM to improve SER performance. However, this study proposes a unique use of CNN upon using it in an MTL framework while utilising the unlabelled data for the auxiliary task to improve SER performance.  For the convenience of the readers, in Table~\ref{table:comparison} we provide a difference of our work with that of the existing literature, which supports the claims we make in this paper.

\section{Proposed Model}
\label{Adver}
We proposed a multi-task learning framework by incorporating semi-supervised adversarial autoencoding using adversarial autoencoders (AAE). An AAE combines a traditional autoencoder and an adversarial network to deliver a surprisingly flexible framework. In AAE, the adversarial part is attached to the latent code z, where the encoder of autoencoder network also acts as the generator of the adversarial network. It enforces the autoencoder to generate a latent representation $z$ by observing the statistical properties of a given prior distribution $p(d)$.

\begin{figure}[!ht]
\centering
\includegraphics[width=0.45\textwidth]{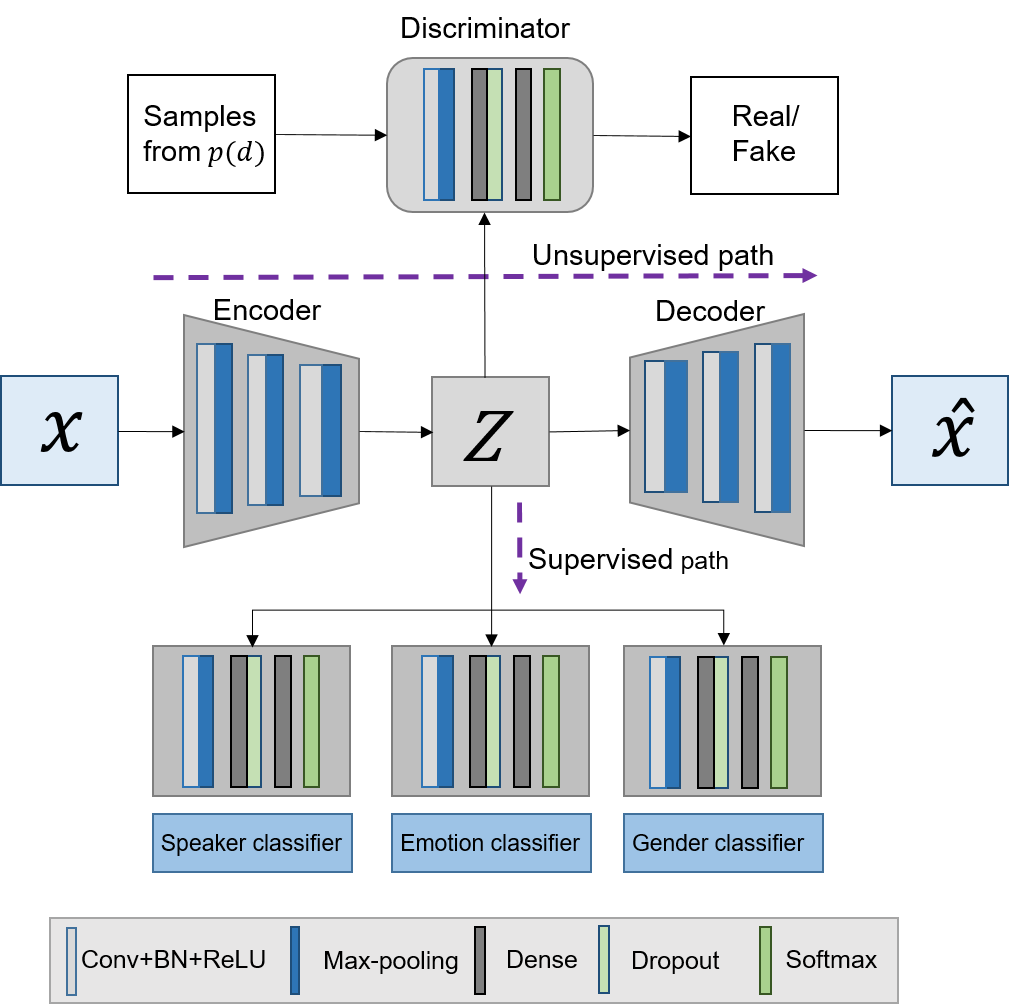}
\caption{Illustration of our proposed multitask framework using a semi-supervised adversarial autoencoder (AAE).}
\label{fig:Model}
\end{figure}

In order to achieve MTL in AAE, we modify it to incorporate three supervised classification networks including emotion, speaker, and gender classification. Fig. \ref{fig:Model} illustrates our proposed semi-supervised multitask learning model, where we highlight the supervised and unsupervised paths. In Equation \eqref{totalloass} we present the multi-task autoencoding loss  $\mathcal{L}_{\text{MTAE}}$ as a function of supervised and unsupervised losses.

\begin{equation}
\label{totalloass}
\mathcal{L}_{\text{MTAE}} = \alpha*\mathcal{L}_{\text{AE}}+\mathcal{L}_{c},
\end{equation}
\begin{equation}
\label{lossc}
\mathcal{L}_{c} = \beta*\mathcal{L}_{\text{E}}+(1-\beta)*\big(\mathcal{L}_{\text{G}}+\mathcal{L}_{\text{S}}\big).
\end{equation}

Here, $\mathcal{L}_{\text{AE}}$ is the reconstruction loss of the autoencoder; $\mathcal{L}_{\text{E}}$, $\mathcal{L}_{\text{G}}$, and $\mathcal{L}_{\text{S}}$  are losses for the emotion, gender, and speaker classification tasks, respectively; $\alpha$ and $\beta$ are the trade-off parameters to control the weight of each loss term.  In addition to the autoencoding network (encoder ($E_{\theta}$) and decoder ($D_{\delta}$)) and the and classifiers ($C_{\phi}$), there is an adversarial network that includes a generator ($E_{\theta}$) and discriminator ($D_{\omega}$). 

For the input data $x$, overall model is trained in three phases: (1) the reconstruction phase; (2) the regularisation phase;  and (3) the classification phase. 
In the reconstruction phase, the autoencoder updates the encoder ($E_{\theta}$) and the decoder ($D_{\delta}$) and minimises the reconstruction error by encoding $x$ into latent representation $z$. The objective function for the autoencoder is defined below:

\begin{equation}
\label{AE}
    \mathcal{L_{\text{AE}}}(x,D_{\delta}(E_{\theta}(x)))=\lVert{x-\hat{x}}\rVert_{2}^{2}
\end{equation}

In \textit{the regularisation phase}, the adversarial network first updates its discriminator ($D_{\omega}$) to distinguish between the samples coming from the prior distribution $p(d)=\mathcal {N}(d; 0,I)$ [real] and that generated using the latent codes $(z)$ [fake] computed by the autoencoder; and then updates its generator ($E_{\theta}$) or encoder .  The objective here is to fool the discriminator ($D_{\omega}$) by learning to encode data that $D_{\omega}$ perceives as real. The update is done by keeping the weight and bias of the discriminator network fixed and by backpropagating the error to $E_{\theta}$ and updating its weight and bias values. 
The objective function for the discriminator ($D_{\omega}$) is defined as: 
\begin{equation}
\begin{aligned}
\text{L}_{\text{disc}} = \underset{\omega}{\text{max}} \big(\mathrm{E}_{d\sim p_{d}}[\log(D_{\omega}(d))]\\ \quad \quad \quad\quad\quad+ \mathrm{E}_{x\sim p_{x}}[\log(1 - D_{\omega}(E_{\theta}(x)))]\big).
\end{aligned}
\end{equation}
Here $p_{x}$ is the data distribution and $p(d)=\mathcal {N}(d; 0,I)$ is the prior multivariate Gaussian distribution.

In the \textit{classification phase}, classifiers ($C_{\phi}$) use the latent code ($z=E_{\theta}(x)$) as input and minimise standard class entropy loss (using predicted values and target vector containing the labels for three tasks) and error is back-propagated through the network to update $E_{\theta}$. The encoder/generator network ($E_{\theta}$) is updated by optimising the following objective function: 
\begin{equation}
\label{enco}
\begin{aligned}
\text{L}_{\text{enc}} = \underset{\theta}{\text{min}} \big(\mathrm{E}_{x\sim p_{x}}[\log(1 - D_{\omega}(E_{\theta}(x)))]\\+\mathrm{E}_{x\sim p_{x}}[\beta\mathcal{L_{\text{AE}}}(x,D_{\delta}(E_{\theta}(x)))]+ \\\mathrm{E}_{x,y \sim p_{X,Y}} [\mathcal{L}_{c}(E_{\theta}(x),y;C_{\phi})]\big).
\end{aligned}
\end{equation}

Unlike the discriminator, the encoder/generator is updated in all three phases. The first term in Equation \eqref{enco} is updated in the regularisation phase, and the second term in the reconstruction phase and the third term is updated in the classification phase. Also, these three phases run in serial: the reconstruction phase runs first followed by the regularisation and classification phase. In this way, the latent code generation, which is an unsupervised process, gets influenced by the supervised classification task and thus facilitates semi-supervised learning. Note here that when using additional auxiliary data with no labels for emotion, loss functions for gender and speaker are only calculated to update the encoder.

\section{Experimental Setup}

\subsection{Datasets}
\label{sec:data}
To evaluate the performance of our proposed model, we use two different datasets: IEMOCAP  and MSP-IMPROV, which are commonly used for emotion classification research \cite{Lotfian+2016,kim2016emotion}. Both datasets have similar labelling schemes and were collected to simulate naturalistic dyadic interactions between actors.  In order to use additional data for the gender and speaker recognition auxiliary tasks, we use Librispeech~\cite{panayotov2015librispeech}, which is a corpus of read English speech, suitable for training and evaluating speech recognition and speaker identification systems. Below we briefly describe these datasets.

\subsubsection{IEMOCAP}
This database contains 12 hours of audiovisual data including audio, video, facial motion information, and textual transcriptions \cite{busso2008iemocap}. The recordings were collected from $10$ professional actors, including five males and five females, during dyadic interactions. This allowed actors to perform spontaneous emotion in contrast to reading text with prototypical emotions \cite{lotfian2017building}.  Each interaction is around five minutes long and segmented into smaller utterances of sentences. For categorical labels, each sentence is annotated by three annotators and the participant. Finally, an utterance is given a label if at least three annotators assigned the same label.  Overall, this corpus contains nine emotions: angry, excited, happy, sad, neutral, disgust, frustrated, fearful, and surprised. For dimensional annotation, two annotators and the participant were asked to label activation and valence on a scale of 1 to 5. Similarly to prior studies \cite{latif2018variational}, we used utterances of four categorical emotions including angry, happy, neutral, and sad in this study by merging ``happy'' and ``excited'' as one emotion class ``happy''. The final dataset includes $5531$ utterances ($1103$ angry,  $1708$ neutral, $1084$ sad, and $1636$ happy).     
\subsubsection{MSP-IMPROV}
The MSP-IMPROV dataset is a multimodal emotional database recorded from $12$ actors performing dyadic interactions \cite{busso2017msp}. The utterances are grouped into six sessions and each session has one male and one female actor similar to IEMOCAP \cite{busso2008iemocap}. The scenarios were carefully designed to promote naturalness, while maintaining control over lexical and emotional contents. The emotional labels were collected through perceptual evaluations using crowdsourcing \cite{burmania2016increasing}. The utterances in this corpus are annotated on four categorical emotions: angry, happy, neutral, sad. To be consistent with previous studies \cite{latif2019direct,gideon2017progressive},  we use  
all utterances with four emotions: anger (792), happy (2644), sad (885), and neutral (3477). For dimensional annotation (i.\,e., activation, and valence), similar to IEMOCAP, these utterances are also annotated on a scale of 1 to 5.

\subsubsection{LibriSpeech}
The LibriSpeech dataset \cite{panayotov2015librispeech} is derived from audiobooks and it contains 1000 hours of English read speech from 2484 speakers. This corpus is commonly used for speech recognition and speaker identification problems \cite{berard2018end,dubey2019transfer}. The training portion of this corpus is split into three subsets, with an approximate recording time of 100, 360 and 500 hours. We used the subset that contains 360 hours of recordings.  These recordings span over 961 speakers. Our selection is motivated by the fact it is obviously larger than the 100-hour subset, which spans over only 251 speakers.  Also, it offers higher recording quality compared to the 500-hour subset \cite{panayotov2015librispeech}. 
From this subset, we randomly select 600 speakers (the rationale for choosing the number of speakers is discussed in Section~\ref{sub_addition_data}).

\subsection{Speech Preprocessing}
We have represented the audio utterances in the form of spectrograms, which is a popular 2D representation widely used for speech emotion recognition \cite{cummins2017image,etienne2018cnn+,albanie2018emotion}. The spectrograms were computed using a short-time Fourier transform (STFT) with an overlapping Hamming window of size 25\,ms with a 10\,ms shift. 
The height of the spectrogram is 128, which represents the frequency range 0--8\,kHz. Due to the varying lengths of the audio samples, the spectrograms vary in width, which poses a problem for the batch processing of the model training. To compensate for this, a context window of 256 frame is applied to create fixed width segments following the procedure used in \cite{chang2017learning,yenigalla2018speech}. Each segmented spectrogram was assigned the emotion label of the corresponding utterance. It is pointed out by previous research that removing silence pauses provides better SER results using deep learning \cite{wang2017learning,satt2017efficient}. One of the reasons is that silence adds no speech information to the training data, especially for deep learning models. Nevertheless, we empirically tested that removing silence pauses offers slightly better performance than retaining them.  In our experiments, we removed silence pauses from the utterances. We trained all models using segmented spectrograms. In order to calculate the utterance level prediction during the testing phase, posterior probabilities of segments of spectrograms for given utterances were averaged. This is a well known strategy used in SER \cite{yenigalla2018speech,latif2019unsupervised} and also in studies on speaker identification \cite{ravanelli2018speaker}.




\subsection{Model Configuration}
Our semi-supervised architecture is illustrated in Figure \ref{fig:Model}. The encoder part of the autoencoder network consists of three convolutional layers.  Each convolutional layer is followed by a pooling layer. These convolutional layers identify emotionally salient regions within the spectrogram and create feature maps. The pooling layer
extract highly relevant features by reducing their dimensions. We use max-pooling layer as it offered better performance compared to average pooling during validation. The encoder/generator part encodes the spectrograms into latent code $z$, which has the  dimension $16\times16\times32$. The size of the latent code was determined using the validation set. Here, we use a multivariate Gaussian distribution ($p(d)=\mathcal {N}(d; 0,I)$) with zero mean and unit standard deviation as prior distribution $p(d)$ that we impose on the latent codes $z$ in the regularisation stage. It helps the AAEs to disentangle important attributes of the input data and makes it suitable for speech emotion classification \cite{parthasarathy2019improving}. In SER, using $\mathcal {N}(d; 0,I)$ as prior helps the autoencoder networks to learn the distribution of emotional structures compared to standard autoencoders as validated with variational autoencoders (VAEs) \cite{latif2018variational} and AAEs \cite{eskimez2018unsupervised,parthasarathy2019improving}.

The model was trained with the batch size of 32, where Stochastic Gradient Descent (SGD) with learning rate of $0.0001$ was used as optimiser. During validation, accuracy was computed at the end of each epoch. If the accuracy of the model did not improve on the validation set after 5 epochs, we restored the model to best epoch and learning rate was halved. This process continued until the learning rate reached below 0.00001.  We applied batch normalisation \cite{ioffe2015batch} 
after each convolutional layer to achieve a stable distribution of activation values throughout the training. The batch normalisation layer was used before the nonlinearity layer. We used a rectified linear unit (ReLU) as non-linear activation function type as it gives us better performance compared to leaky ReLU and hyperbolic tangent during validation.
The decoder block has the same structure as the encoder/generator except that the convolutional layers are replaced with transposed convolution layers. 

The latent code $z$ was fed to the classifiers, which has four components: (1) convolutional layer, (2) max-pooling layer, (3) dense layers, and (4) softmax layer. We used one convolutional layer followed by max-pooling in each classifier to capture features related to the classification tasks. After each max-pooling layer, we used dense layers followed by a softmax layer to provide prediction. We used two dense layers in each classifier and used a dropout layer, with a dropout rate of 0.3,
between them to avoid overfitting. The discriminator of the AAE had a similar architecture to the classifiers, which consists of one convolutional layer followed by a max-pooling layer, and two dense layers followed by a softmax layer. 

We performed the step-by-step training of all models. We randomly initialised the model and trained first on Librispeech dataset for speaker and gender classification only. The weights learnt in this stage were used to initialise the autoencoder network when emotional data is fed to the model.



\section{Experiments and Evaluations}
In this study, we evaluated the performance of the proposed framework using 10-fold cross-validation and leave-one-speaker-out validation to compare with multiple studies. For 10-fold cross-validation, we followed the strategies used in~\cite{gideon2017progressive}. We created the ten folds based on speaker ID so that each fold has all speakers. This allows us to use speaker identification as secondary task.  In each step of the validation, one fold was used as validation set for parameter selection, eight folds were used for training, and the remaining fold was used for testing (same as used in~\cite{gideon2017progressive} ). 

The 10-fold cross-validation scheme does not allow for speaker-independent testing as such since each fold has data from all speakers. To perform speaker-independent testing we used the leave-one-speaker-out cross-validation scheme commonly used in the literature \cite{xia2017multi,latif2018variational}. This ensures that the speakers are independent in each fold.  However, speaker-independent testing limits the use of speaker identification as secondary task. Therefore, we performed speaker verification in this scheme.  We consider the d-vector framework \cite{variani2014deep}  for speaker verification, which uses the output of the last hidden layer as speaker representation. During training, the speaker classifier is trained to classify speakers on training data and the evaluation phase involves the extraction of a d-vector from the test utterance using the trained speaker classifier. Then cosine distance is computed between the d-vectors of the test and the claimed speaker. The standard test set of the Librispeech data \cite{panayotov2015librispeech} consisting of 40 speakers along with the test sets of IEMOCAP or MSP-IMPROV were used for enrolment and verification purposes. Equal Error Rate (EER) is used as a measure of performance in the speaker verification system. For SER, we used weighted accuracy (WA) and unweighted accuracy (UA) as comparison measures due to their widely accepted use in studies on speech emotion recognition. For each model used in this work, we repeated the evaluation ten times and calculated the mean and standard deviation of WA and UA.


For evaluations, we considered two types of emotional labels: categorical and dimensional. For categorical emotions, we used four emotions including anger, sad, happy, and neutral. For dimensional emotions, we used two different configurations. Firstly, we manually clustered continuous values of dimensional emotions into three levels. We interpreted activation as low, medium, high, and valence as negative, neutral, positive, as used in \cite{xia2017multi,wollmer2010context}. Both, the IEMOCAP and MSP-IMPROV databases are annotated for activation and valence using integral values in the range $1$ to $5$. Table \ref{table:3level} shows the range of three clusters for activation and valence for both datasets, which has also been adopted from~\cite{xia2017multi,wollmer2010context}.
\begin{table}[ht]
\scriptsize
\centering
\caption{Three Levels of Mapping Rules for IEMOCAP and MSP-IMPROV.}
\begin{tabular}{|m{2.1cm}|m{1.6cm}|m{1.8cm}|m{1.5cm}|}
\hline
\textbf{Corpus}
&\textbf{Low/Negative}
&\textbf{Medium/Neutral}
&\textbf{High/Positive}
\\ \hline
\begin{tabular}[c]{@{}l@{}}IEMOCAP\end{tabular}
&\begin{tabular}[c]{@{}l@{}}[1,2]\end{tabular}
&\begin{tabular}[c]{@{}l@{}}(2,3.5]\end{tabular}
&\begin{tabular}[c]{@{}l@{}}(3.5,5]\end{tabular}
\\\hline
\begin{tabular}[c]{@{}l@{}}MSP-IMPROV\end{tabular}
&\begin{tabular}[c]{@{}l@{}}[1,2.5]\end{tabular}
&\begin{tabular}[c]{@{}l@{}}(2.5,3.5]\end{tabular}
&\begin{tabular}[c]{@{}l@{}}(3.5,5]\end{tabular}
\\\hline
\end{tabular}
\centering
\label{table:3level}
\end{table}


\begin{figure}[!ht]
\centering
\includegraphics[ width=0.45\textwidth]{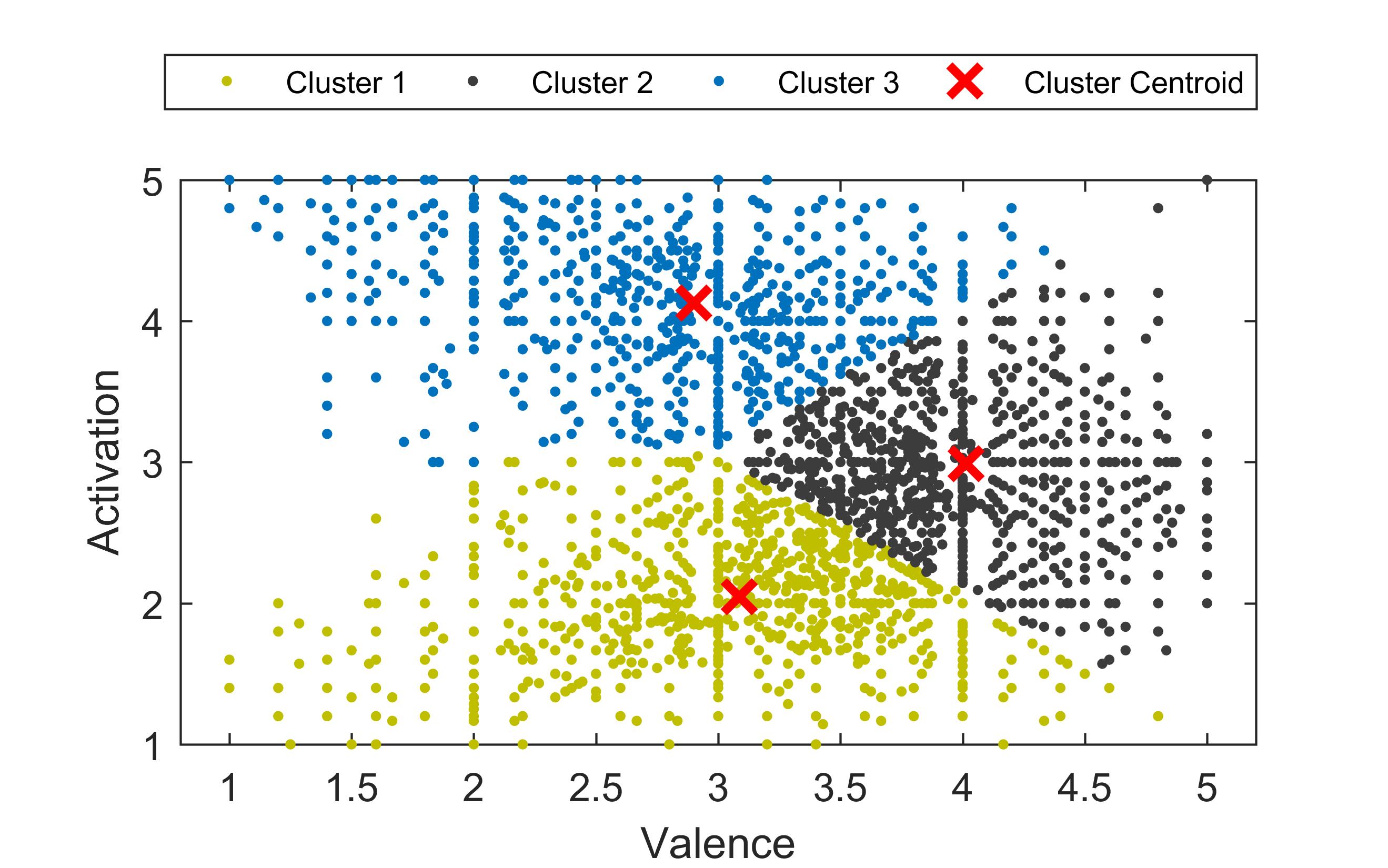}
\caption{ Three clusters of the MSP-IMPROV data in the valence-activation space.}
\label{fig:clus}
\end{figure}

\begin{figure*}[!t]%
\centering
\begin{subfigure}{0.49\linewidth}
\includegraphics[trim=0.1cm 0.1cm 0.2cm 0.5cm,clip=true,width=\linewidth]{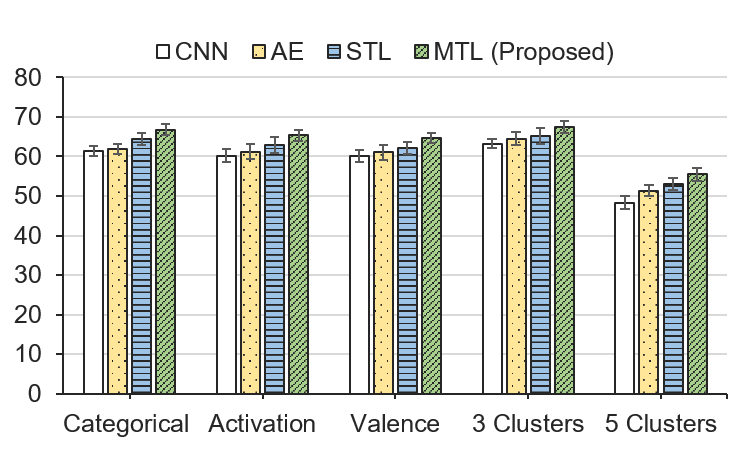}%
\captionsetup{justification=centering}
\caption{IEMOCAP}%
\label{IEMOCAP}%
\end{subfigure}\hfill%
\begin{subfigure}{0.49\linewidth}
\includegraphics[trim=0.0cm 0.1cm 0.2cm 0.5cm,clip=true,width=\linewidth]{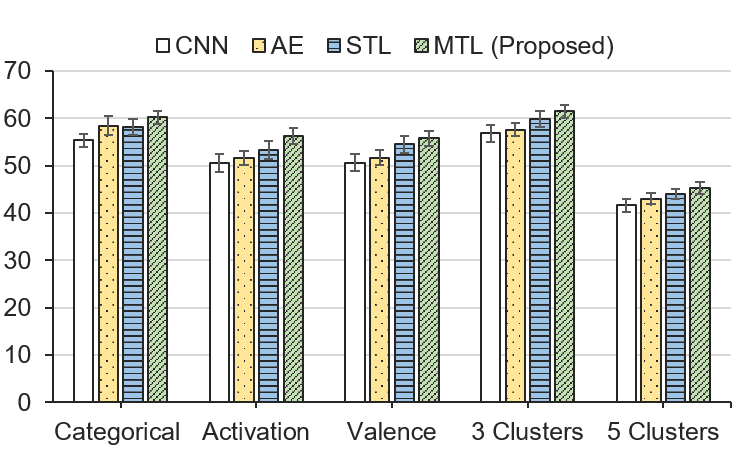}%
\captionsetup{justification=centering}
\caption{MSP-IMPROV} %
\label{MSP-IMPROV}%
\end{subfigure}%
\caption{Benchmarking results of the proposed multi-task model (MTL) against a single task implementation of the same model (STL), single task implementation by CNN, and a single-task semi-supervised implementation of an autoencoder (AE) using leave-one-speaker-out scheme}.
\label{fig:benchmark}
\end{figure*}

Secondly, we considered dimensional emotion representation using the valence-activation space, which combines the information of both activation and valence. Previous work has shown that the combination of these two dimensions provides richer emotional information in contrast to using valence and activation separately \cite{lee2009modeling}. To be consistent with previous studies \cite{mariooryad2013exploring,xia2017multi,metallinou2012context}, we validated our model on the joint classification of the valence-activation space by building three and five clusters. The cluster midpoints in the valence-activation space were determined by applying $K$-means clustering on the dimensional annotation values of the respective datasets (i.\,e., IEMOCAP and MSP-IMPROV). A label was assigned to each utterance by choosing the cluster label that minimised the Euclidean distance between the utterance and the cluster centroid. This is highlighted in Figure \ref{fig:clus}.



\subsection{Benchmarking Results}
We start our evaluation by benchmarking the performance for multi-task over single-task. We implemented a single-task learning (STL) version of our proposed model for a fair comparison. Also, to expand our comparison, we implemented a supervised convolutional neural network (CNN) and an autoencoder (AE) based semi-supervised framework, both using single task learning. We compared the performances of these models for both categorical and dimensional emotions. 

As mentioned, while investigating categorical emotion, we use it as the primary task and speaker verification and gender identification as the auxiliary tasks. Then, we use dimensional emotions as the primary task in two ways: grouping continuous values into three levels; and clustering valence-arousal space into three different groups as discussed before.
For both ways, the secondary tasks are gender classification and speaker verification.


We trained the single task implementation of our model and the autoencoder in a semi-supervised way for the emotion recognition task. The overall loss function was optimised by tuning the values of $\alpha$ and $\beta$ in Equation \eqref{totalloass} and \eqref{lossc} to maximise the system performance and minimise the reconstruction loss, $\mathcal{L}_{\text{AE}}$. Here, Equation \eqref{lossc} contains only first term ($\mathcal{L}_{c} = \beta*\mathcal{L}_{\text{E}}$) as we are only using emotion classifier. We evaluated the model for different values of $\alpha$ and and $\beta$ ranging from $0.1$ to $1.0$ on the validation set to select the best value. For the IEMOCAP data, we achieve the best performance for the AE using $\alpha=0.3$ and $\beta=0.7$, and for our AAE based models, we achieve the best results for $\alpha=0.4$ and $\beta=0.6$.  

For two configurations of dimensional emotions, we also identify $\alpha$ and $\beta$ using the validation set and use it for the test set. We achieve the best performance for the AE using $\alpha=0.6$ and $\beta=0.4$, and for our AAE based models, we achieve the best results using $\alpha=0.3$ and $\beta=0.7$. %

The comparisons of results are all summarised in Fig.~\ref{fig:benchmark}. We observe that, our proposed MTL framework performs better than the STL implementation of our model---the supervised CNN, and the autoencoder.  We note this for both categorical and dimensional classification of emotion and for both the IEMOCAP and MSP-IMPROV datasets.

 \subsection{Comparison with Previous Studies}
To further extend our comparison scope, in this section, we include results published in recent studies. Note that for IEMOCAP and MSP-IMPROV, there are no standardised training and testing splits to evaluate the results. However, we observe that most of the related studies have used either a 10-fold or a leave-one-speaker-out validation strategy. We therefore implement these schemes and present the comparison results in Table \ref{table:comparision}. These are, however, accordingly, to be interpreted with the necessary care and merely serve as indication.

\begin{table}[!h]
\centering
\scriptsize
\caption{Comparison of results (UA \%) of our proposed method with those of recent studies using categorical emotions.}
\begin{tabular}{|l|l|l|}
\hline
\multicolumn{3}{|c|}{{\bf 10-fold cross validation results}} \\ \hline 
\multirow{1}{*}{Model} & \multicolumn{1}{l|}{\tiny{IEMOCAP}} & \multicolumn{1}{l|}{\tiny{MSP-IMPROV}} \\ \hline  
ProgNet (Transfer Learning) \cite{gideon2017progressive}  &  65.7  $\pm$1.8       &       60.5 $\pm$2.1           \\ \hline
CNN (Muli-task implementation)       &   65.6 $\pm$2.0                    & 59.5 $\pm$2.4              \\ \hline
Semi-supervised AE (Muli-task implementation)         &   66.4 $\pm$1.6                   & 60.2 $\pm$2.3              \\ \hline
Semi-supervised AAE (Proposed)       & {\bf 68.8$\pm$1.2}                        &  {\bf 63.6$\pm$1.7}             \\\hline
\multicolumn{3}{|c|}{{\bf leave-one-speaker-out}} \\ \hline 
DBN (Multi-task) \cite{xia2017multi}&     62.2                   &  --            \\ \hline
CNN (Multi-task) \cite{neumann2019improving}&     59.54                  &  --            \\ \hline
Semi-supervised AAE (proposed) &   {\bf 66.7$\pm$1.4}         &  {\bf 60.3$\pm$1.1}    \\ \hlineB{2.5}
CVAE-LSTM (Single Task) \cite{Latif2018v} &   62.8              & --                          \\ \hline
CNN  (Single Task)  \cite{ma2018emotion}      &64.2                 &  --\\ \hline
Proposed  (Single Task)       &64.5$\pm1.5$                 & 58.1$\pm1.7$\\ \hline

\end{tabular}
\label{table:comparision}
\end{table}

 For 10-fold cross validation, we followed the evaluation scheme used in \cite{gideon2017progressive}. In~\cite{gideon2017progressive}, authors used progressive neural network and transfer learning (TL) to transfer knowledge from gender and speaker identification to improve the SER performance. Compared to this study, we are achieving better results by exploiting speaker and gender identification as auxiliary tasks within our multi-task learning framework. This shows that, transferring knowledge using auxiliary tasks in MTL can provide more useful information to improve SER performance. 
 In Table \ref{table:comparision}, we also report the performance of a CNN and an AE when implemented for multi-task learning. We implemented a multitask CNN with two convolutional layers shared with the classification networks following the technique used in \cite{li2014heterogeneous,chang2017learning}. The classification networks consisted of one convolutional layer, two dense layers, and one softmax layer. The AE model had a similar architecture (i.\,e., hidden units, layers, and model parameters) as our AAE based model---just without the discriminator. 

In order to evaluate the proposed model for speaker-independent SER, we used leave-one-speaker-out training with five-fold cross-validation.  As speaker independent scheme limits speaker identification as an auxiliary task, therefore, we performed speaker verification. We compare our results with recent studies \cite{Latif2018v,ma2018emotion,xia2017multi} using speaker-independent SER.  In \cite{xia2017multi}, the authors used a multi-task DBN for SER and showed the improved results compared to STL. In \cite{neumann2019improving} the authors used multi-task CNN and utilised additional unlabelled data in an unsupervised way to improve SER performance. Similar to these studies, we also consider MTL framework and achieved better performance than their approaches as reported in Table \ref{table:comparision}.  The authors in \cite{xia2017multi} and \cite{neumann2019improving} utilised other emotional attributes as auxiliary tasks which limit the use of additional data. 

While using the leave-one-speaker-out training with five-fold cross-validation, our proposed model in STL setting also provides better results compared with the other recent studies \cite{Latif2018v,ma2018emotion} using STL. Note that these studies did not report any result for the MSP-IMPROV dataset.  Also note that, due to the difference in the activation and valence classification strategies, we could not present the results of \cite{chang2017learning} in Table \ref{table:comparision}.



\subsection{Cross-Corpus Results}
To verify the generalisability of the proposed model, we also perform a cross-corpus analysis. In this scenario, we trained models using IEMOCAP, and testing is performed on the MSP-IMPROV set. We selected IEMOCAP as training data since it is more balanced and also for good comparison with recent studies, as these studies used a similar scheme \cite{bao2019cyclegan,sahu2018enhancing,neumann2019improving}. We used 30\,\% of the MSP-IMPROV data for parameter selection and 70\,\% as testing data. Here, we used gender classification and speaker verification as an auxiliary task, as speakers in both datasets are different.

We compared our results with other studies on cross-corpus SER. For example, Neumann et al.\  \cite{neumann2019improving} utilised the representations learnt by autoencoder from unlabelled data fed into a CNN-based classifier. They used the Librispeech and Tedlium (release 2) \cite{rousseau2014enhancing} datasets as unlabelled data, and were able to improve the performance for cross-corpus SER. Our proposed model provides better results compared to this study by using additional data for auxiliary task. In \cite{bao2019cyclegan}, the authors used Cycle consistent adversarial networks, i.\,e., the (CycleGAN)-based method to transfer feature vectors extracted from a large unlabelled speech corpus into synthetic features representing the given target emotions. They used Tedlium (release 2) as unlabelled data to generate synthetic data and used this data to augment the classifier. Similarly, Sahu et al.\ \cite{sahu2018enhancing} applied generated samples by GANs as additional data to train the classifiers for cross-corpus SER. Both of these studies \cite{bao2019cyclegan,sahu2018enhancing} used additional data to augment the classifiers for cross-corpus SER. However, our approach is different as we are using additional data for auxiliary tasks and achieving similar results without augmenting the system with synthetic data. We compare our results with those of these studies and the comparisons are presented in Table \ref{cross}. The results show that we achieve competitive accuracy attesting the generalisation ability of the proposed model.

\begin{table}[!ht]
\centering
\caption{Cross-corpus evaluation results for emotion recognition.}
\begin{tabular}{|l|l|}
\hline
Model                                         & UA (\%)\\ \hline
\begin{tabular}[c]{@{}l@{}}Attentive CNN \cite{neumann2019improving}\end{tabular} &45.76    \\ \hline
\begin{tabular}[c]{@{}l@{}}Conditional-GAN \cite{sahu2018enhancing}\end{tabular} &45.40    \\ \hline
\begin{tabular}[c]{@{}l@{}}CycleGAN-DNN \cite{bao2019cyclegan}\end{tabular} & 46.52$\pm$0.43 \\ \hline
\begin{tabular}[c]{@{}l@{}}Proposed\end{tabular} & 46.41$\pm$0.32   \\ \hline
\end{tabular}
\label{cross}
\end{table}

\section{Analysis and Discussion}
The experimental results clearly show that the proposed semi-supervised multi-task framework offers an improved performance in speech emotion recognition compared to previous studies.
In this section, we focus on three aspects of our proposed model: (1) we elaborate on the impact of a secondary task on improving the performance of the primary task; (2) we quantify the impact of using additional data; and (3) we quantify the impact of tuning the trade-off parameters. All the results in this section are computed using speaker-independent evaluation. 

\begin{table}[!ht]
\scriptsize
\centering
\caption{Impact of auxiliary tasks on categorical emotions.}
\begin{tabular}{|c|c|c|c|c|}
\hline
\multirow{3}{*}{Secondary Tasks} & \multicolumn{4}{c|}{Primary task: categorical emotion}          \\ \cline{2-5} 
 & \multicolumn{2}{c|}{IEMOCAP} & \multicolumn{2}{c|}{MSP-IMPROV} \\ \cline{2-5} 
 & WA            & UA           & WA             & UA             \\ \hline
                                                    
Gender &$67.5$$\pm$1.5&$66.1$$\pm$1.7&$60.1$$\pm$1.2&$59.3$$\pm$1.1\\ \hline
Speaker  &$67.2$$\pm$1.3&$65.9$$\pm$1.6&$60.5$$\pm$1.2&$59.1$$\pm$1.5  \\ \hline
Both&{\bf68.5$\pm$1.2}&{\bf66.7$\pm$1.4}&{\bf 62.5$\pm$1.4}&{\bf 60.2$\pm$1.2}  \\ \hline
\end{tabular}
\label{MTL_C}
\end{table}

\begin{table}[!ht]
\centering
\scriptsize
\caption{Impact of auxiliary tasks on dimensional emotions.}
\begin{tabular}{|l|c|c|c|c|}
\hline
\multirow{3}{*}{\scriptsize{Secondary Tasks}} & \multicolumn{4}{c|}{Primary Tasks: dimensional emotion}                             \\ \cline{2-5} 
&\multicolumn{4}{c|}{Individual levels of activation and valence}                             \\ \cline{2-5}
 & \multicolumn{2}{c|}{IEMOCAP} & \multicolumn{2}{c|}{MSP-IMPROV} \\ \cline{2-5} 
 & \scriptsize{Activation}   & \scriptsize{Valence}      & \scriptsize{Activation}     & \scriptsize{Valence}        \\ \hline
Gender  &  62.6$\pm$1.3   &   61.4$\pm$1.2   &   53.7$\pm$2.1        &  53.4 $\pm$1.8           \\ \hline
Speaker    &  63.7  $\pm$1.2 &  60.8$\pm$0.8 &   52.6  $\pm$2.5         &   52.6$\pm$2.3            \\ \hline
Both     & {\bf 64.5$\pm$1.5}   &  {\bf 62.2$\pm$1.0}     &    {\bf 54.6$\pm$1.4}      &    {\bf 55.4$\pm$1.6}            \\ \hline
\multirow{2}{*}{\scriptsize{Secondary Tasks}} & \multicolumn{4}{c|}{Joint activation-valence space}                             \\ \cline{2-5} 
          & \scriptsize{3 Clusters}    & \scriptsize{5 Clusters}   & \scriptsize{3 Clusters}     & \scriptsize{5 Clusters}     \\ \hline

Gender     &65.7$\pm$1.3          &  53.2 $\pm$0.8       &  60.1$\pm$1.5                    &  43.6$\pm$1.5      \\ \hline
Speaker     & 64.8 $\pm$1.5     &  54.2  $\pm$1.2       &   61.4$\pm$1.0      &     43.1$\pm$1.2        \\ \hline
Both         &  {\bf 65.1$\pm$0.9}         &   {\bf 55.1$\pm$1.4}         &  {\bf 62.1$\pm$1.8}             & {\bf 44.3$\pm$1.3}           \\ \hline
\end{tabular}
\label{dimen}
\end{table}

\begin{figure*}[!ht]%
\centering
\begin{subfigure}{0.25\linewidth}
\includegraphics[trim=0.45cm 0.5cm 0.2cm 0.4cm,clip=true,width=\linewidth]{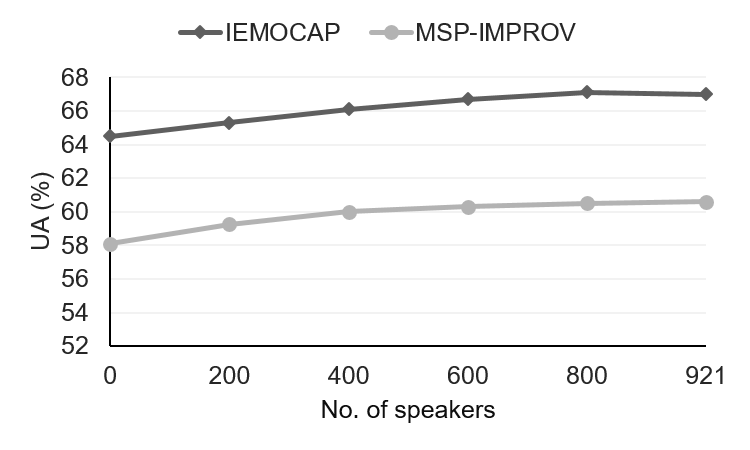}%
\captionsetup{justification=centering}
\caption{emotion recognition} %
\label{Emotion_C}%
\end{subfigure}%
\begin{subfigure}{0.25\linewidth}
\includegraphics[trim=0.45cm 0.5cm 0.2cm 0.4cm,clip=true,width=\linewidth]{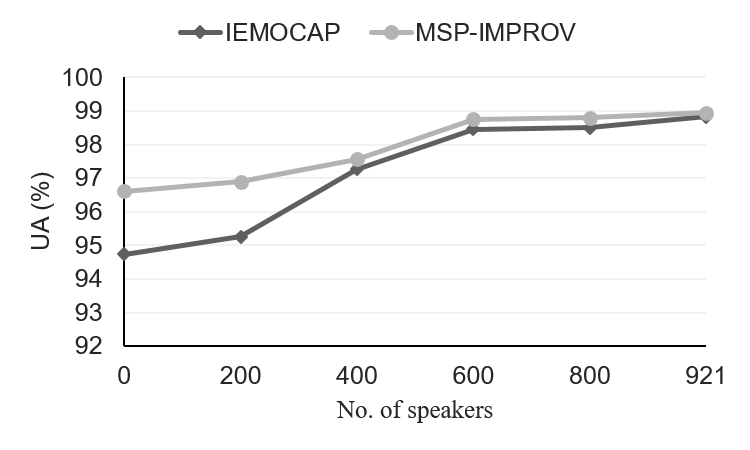}%
\captionsetup{justification=centering}
\caption{gender classification}%
\label{Gender_C}%
\end{subfigure}
\begin{subfigure}{0.25\linewidth}
\includegraphics[trim=0.45cm 0.5cm 0.2cm 0.4cm,clip=true,width=\linewidth]{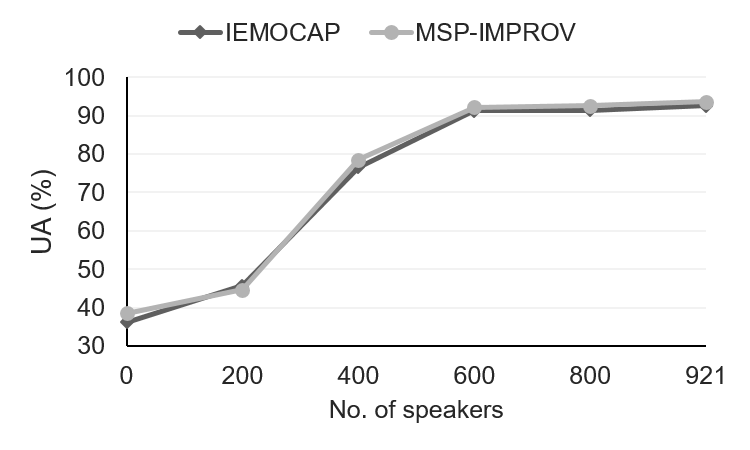}%
\captionsetup{justification=centering}
\caption{speaker identification} %
\label{Speaker_C}%
\end{subfigure}%
\begin{subfigure}{0.25\linewidth}
\includegraphics[trim=0.45cm 0.5cm 0.2cm 0.4cm,clip=true,width=\linewidth]{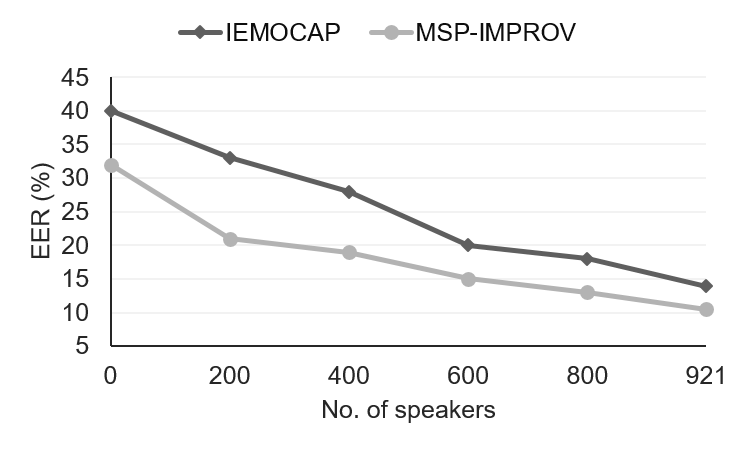}%
\captionsetup{justification=centering}
\caption{speaker verification } %
\label{Speaker_V}%
\end{subfigure}%
\caption{Impact of additional data injection on gender, speaker and emotion recognition. }
\label{fig:dataeffect}
\end{figure*}

\subsection{Impact of Secondary Tasks on Primary Task}
We consider four categorical emotions from both the IEMOCAP and MSP-IMPROV datasets as described in Section \ref{sec:data}. We also use dimensional emotions in our experiments. 
In Table \ref{MTL_C} and Table \ref{dimen}, we present the results of using the auxiliary tasks separately and jointly to improve the performance of the primary task for categorical and dimensional emotions, respectively. We observe that, while using the auxiliary tasks individually, our model offers similar performance improvement. However, when using the auxiliary tasks jointly, our model offers the highest accuracy for the primary task across both categorical and dimensional emotion representations. Intuitively, jointly learning a representation for emotions with speaker and gender helps to uncover the common high-level discriminative representations, which leads to the performance improvements in the SER system.  


\subsection{Impact of using Additional Data}
\label{sub_addition_data}
For the auxiliary tasks of speaker and gender recognition, we use additional data that is not labelled for emotion and show that, when the MTL model is trained with additional data from auxiliary tasks, the performance on the emotion recognition task for both datasets. To further show how performance improves while increasing the amount of data, we trained our model by varying the amount of data for auxiliary tasks. Note that we use the LibriSpeech dataset to introduce additional speakers so, in order to increase the amount of data, we increase the number of speakers. 

Fig.\ \ref{Emotion_C} shows the effect of varying the amount of additional data on the UA (\%) of categorical emotion classification using both datasets. 
We observe that up to 600 speakers, the performance improvement is quite strong,
however, beyond that we observe a plateauing effect. This is an important observation as it can guide researchers to select a possible operating point when using our suggested method. 

To get some further insight into the above improvement, we plot the improvement in auxiliary tasks with the increase of data. 
Fig.~\ref{Gender_C}, Fig.~\ref{Speaker_C}, Fig.~\ref{Speaker_V} summarise the results. Here, the performance is calculated using our model in the single task learning mode. We plot the results as we increase the number of speakers. We notice a similar trend as we observe in Fig.\ \ref{Emotion_C}. After 600 speakers, improvement in secondary tasks sees a plateau effect.  Here, we also plot the speaker verification EER (\%) with the increase of number of speakers by using 20 utterances of speakers for enrolment and remaining for evaluation. The performance of speaker verification also improves with the increase of speaker data. Therefore, summarising Fig.\ \ref{fig:dataeffect}, it can be noted that improvement in the auxiliary tasks while adding additional data eventually helps to improve the performance of the primary task, which cements the contribution of this paper in proposing a multi-task semi-supervised framework for SER. Intuitively, through the feed of additional data for the auxiliary tasks, a better representation of the intrinsic properties of speech is achieved, which eventually improves the performance of SER. 


\begin{figure}[!ht]%
\centering
\begin{subfigure}{\linewidth}
\includegraphics[trim=0.4cm 0.5cm 0.2cm 0.1cm,clip=true,width=0.5\linewidth]{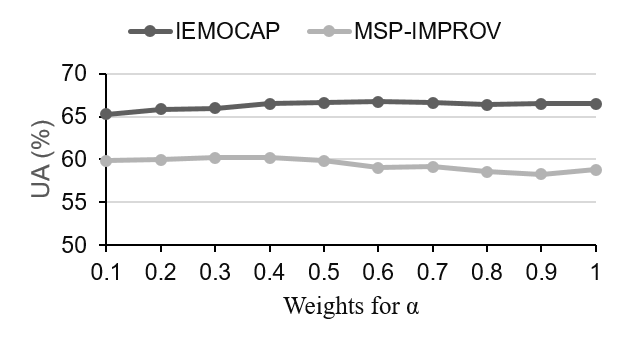}
\includegraphics[trim=0.4cm 0.5cm 0.2cm 0.1cm,clip=true,width=0.5\linewidth]{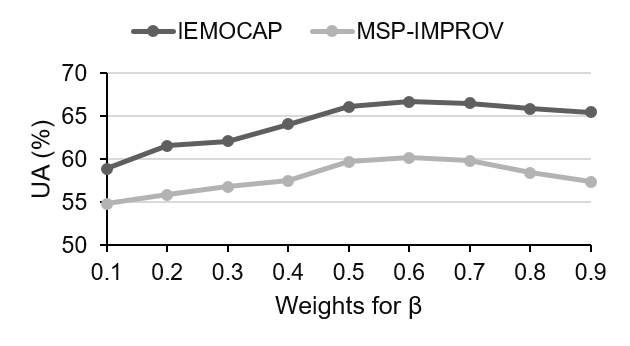}%
\captionsetup{justification=centering}
\caption{\scriptsize{categorical emotion classification}}%
\label{cat}%
\end{subfigure}
\vskip\baselineskip
\begin{subfigure}{\linewidth}
\includegraphics[trim=0.4cm 0.5cm 0.2cm 0.1cm,clip=true,width=0.5\linewidth]{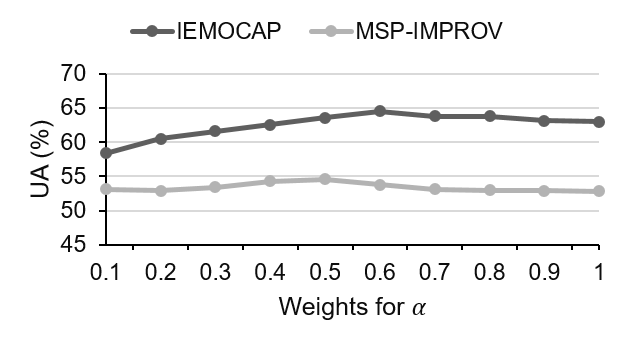}%
\includegraphics[trim=0.4cm 0.5cm 0.2cm 0.1cm,clip=true,width=0.5\linewidth]{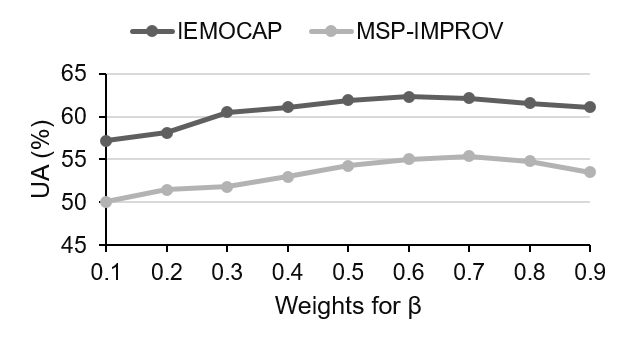}%
\captionsetup{justification=centering}
\caption{\scriptsize{activation classification}}%
\label{act}
\end{subfigure}%
\vskip\baselineskip
\begin{subfigure}{\linewidth}
\includegraphics[trim=0.4cm 0.5cm 0.2cm 0.1cm,clip=true,width=0.5\linewidth]{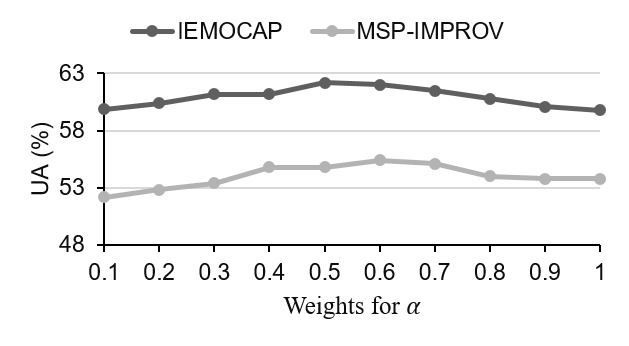}%
\includegraphics[trim=0.4cm 0.5cm 0.2cm 0.1cm,clip=true,width=0.5\linewidth]{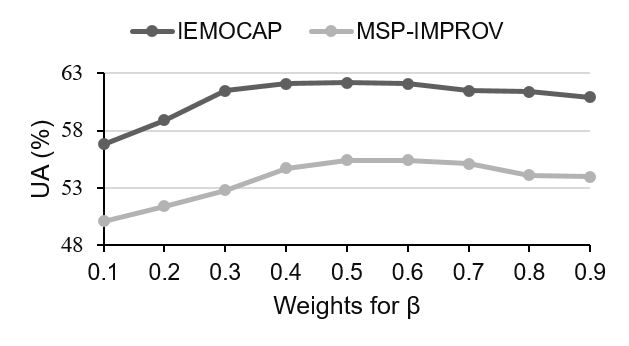}%
\captionsetup{justification=centering}
\caption{\scriptsize{valence classification}}%
\label{val}
\end{subfigure}%

\caption{Impact of varying the weights of $\alpha$ and $\beta$ on the performance of categorical (\ref{cat}), activation (\ref{act}), and valence (\ref{val}) classification on both, the IEMOCAP and MSP-IMPROV datasets. }
\label{fig:alpha}
\end{figure}

\subsection{Impact of Tuning Trade-off Parameters}
In this section, we investigate the impact of the trade-off parameters $\alpha$ (Eq~\eqref{totalloass}), and $\beta$ (Eq~\eqref{lossc}), which are the weights of unsupervised, and supervised primary and secondary tasks, on the performance (UA) of the system.

Fig~\ref{fig:alpha} shows the impact of changing the weights $\alpha$ and $\beta$ on UA (\%) for categorical, activation, and valence classification. In the first experiment, we keep the value of $\beta$ fixed at $0.5$. This assigns equal weights to both the primary and secondary tasks, but vary the weights (0.1 to 1.0) for the  unsupervised task. It can be seen from Fig. \ref{fig:alpha} that very low and very high weights of $\alpha$ hurt the performance of the system. However, $\alpha$ with values ranging from $0.4-0.6$ gave better results for both datasets. 
This shows that controlling the weights of the unsupervised task through $\alpha$ can improve the performance of the system, however, a suitable range for $\alpha$ needs to be identified that offers the best performance.

Further, Fig.~\ref{fig:alpha} also illustrates the relationship of $\beta$ and UA (\%). To highlight this, we select $\alpha=1$ and vary the weights of $\beta$ (0.1 to 0.9), which controls the weights for both the primary and secondary tasks classification losses (see Equation \eqref{lossc}). 
It can be noted from Figs.\ ~\ref{cat}, \ref{act}, and \ref{val} that a very high value of $\beta$ (i.\,e., the frameworks essentially become single task) gives poor performance. However, we also note that too much significance given to auxiliary tasks also diminishes the performance as a very low value of $\beta$ gives poor performance. For both the IEMOCAP and MSP-IMPROV datasets, the system performed better with values of $\beta$ in the range of $0.4-0.7$.

\section{Conclusions}
In contrast with previous studies, this article proposes semi-supervised multi-task learning using adversarial autoencoders for speech emotion recognition (SER). Specifically, we put considerable emphasis on a novel technique of utilising unlabelled data for auxiliary tasks through the proposed multi-task semi-supervised learning model to improve the accuracy of the primary task. We evaluated our proposed model using the popular IEMOCAP and MSP-IMPROV emotion datasets, and demonstrated that it performs notably  better than (1) the comparable state-of-the-art studies in SER that use similar methodology and/or implementation strategies; (2) supervised single- and multi-task methods based on CNN, and (3) single- and multi-task semi-supervised autoencoders. We observe this for categorical and dimensional emotion classifications, and cross-corpus SER. Our proposed approach can overcome the challenge of limited data availability of emotion datasets, which is a significant contribution
towards developing a robust machine learning model for SER.

Our analysis shows that (1) improvement of the auxiliary tasks through the injection of additional data predominantly drives the improvement of the primary task, (2) a combined effort of auxiliary task is better for improving the accuracy of the primary task, than  using them individually, (3) for the IEMOCAP and MSP-IMPROV datasets, it is possible to reasonably determine an operating point in terms of how much additional data for the auxiliary task is sufficient, (4) it is important to control the weight of loss function of the unsupervised task in the proposed semi-supervised MTL setting to improve the accuracy of SER, and (5) it is important to control the weight of the loss functions of the primary and secondary tasks to achieve the best possible accuracy for SER. 


Future work should further focus on the tighter coupling between the generation of data and modelling a richer selection of speaker states and traits simultaneously aiming at `holistic' speaker analysis \cite{zhang2017multi}. In addition, it appears highly attractive to integrate reinforcement learning into such a framework given a real-life usage such as in a dialogue manager. Likewise, semi-supervised and unsupervised aspects can be benefited by reinforced information.

\end{document}